\documentclass[11pt,twoside]{article}
\input{epsf.sty}
\usepackage{mathrsfs}
\usepackage{amsmath}
\usepackage{amssymb}
\usepackage{graphicx}
\usepackage{subfigure}
\newtheorem{proposition}{Proposition}

\title{A three-component Camassa-Holm (3CH) system with cubic nonlinearity and peakons}
\author{Baoqiang Xia$^{1}$\footnote{E-mail address: xiabaoqiang@126.com},
~~Ruguang Zhou$^{1}$\footnote{E-mail address: zhouruguang@jsnu.edu.cn},
~~Zhijun Qiao$^{2}$\footnote{E-mail address: qiao@utpa.edu}
\\
$^1$School of Mathematics and Statistics, Jiangsu Normal
University,\\
 Xuzhou, Jiangsu 221116, P. R. China
\\ $^{2}$Department of Mathematics, University of Texas-Pan American, \\Edinburg, Texas 78541, USA}

\date{}
\begin{document}
\maketitle
\begin{abstract}
In this paper, we propose a three-component  Camassa-Holm (3CH) system with cubic nonlinearity and peakons. The 3CH model is proven integrable in the sense of Lax pair, Hamiltonian structure, and conservation laws. We show that this system admits peaked soliton (peakon) and multi-peakon solutions. Additionally, reductions of the 3CH system are investigated so that a new integrable perturbed CH equation with cubic nonlinearity is generated to possess peakon solutions.


\end{abstract}

\section{ Introduction}
In the past two decades, the Camassa-Holm (CH) equation \cite{CH}
\begin{eqnarray}
m_t-2m u_x-m_xu=0, \quad m=u-u_{xx}+k,
\label{CH}
\end{eqnarray}
with $k$ being an arbitrary constant, derived by Camassa and Holm \cite{CH} as a shallow water
wave model, has attracted much attention in the theory of soliton and integrable system.
The CH equation was first included in the work of Fuchssteiner and Fokas on hereditary symmetries as a very special case \cite{FF1}.
Since the work of Camassa and Holm \cite{CH}, more diverse studies on this equation have remarkably been developed \cite{CH2}-\cite{CGI}.
The most interesting feature of the CH equation (\ref{CH}) is admitting peaked
soliton (peakon) solutions in the case of $k=0$. A peakon is a kind of weak
solution in some Sobolev space with corner at its crest.
The stability and interaction of peakons were discussed in several references \cite{CS1}-\cite{JR}.

As extension of the CH peakon equation, other integrable peakon models have also been found,
such as the Degasperis-Procesi (DP) equation \cite{DP1,DP2}
\begin{eqnarray}
m_t+3m u_x+m_xu=0, \quad m=u-u_{xx},
\label{DP}
\end{eqnarray}
the cubic nonlinear peakon equation \cite{OR,Fo,Fu,Q1}
\begin{eqnarray}
 m_t=\left[ m(u^2-u^2_x)\right]_x, \quad  m=u-u_{xx},\label{mCH}
\end{eqnarray}
and the Novikov's cubic nonlinear equation \cite{NV1,HW1}
\begin{eqnarray}
 m_t=u^2m_x+3uu_xm, \quad  m=u-u_{xx}.\label{cCHN}
\end{eqnarray}
Then, a naturally interesting theme is to study integrable multi-component peakon equations.
For example, in \cite{OR,CLZ,Fa,HI} the authors proposed the two-component generalizations of the CH equation.
In \cite{SQQ}-\cite{GX1}, the two-component extensions of the cubic nonlinear equations (\ref{mCH}) and (\ref{cCHN}) were investigated, and while in \cite{FL,GX,QF}, the three-component extensions of CH equation are derived.

In this paper, we propose the following three-component system
\begin{eqnarray}
\left\{\begin{split}
m_{11,t}=&\frac{1}{2}[m_{11}(u_{11}^2-u_{11,x}^2+u_{12}u_{21}-u_{12,x}u_{21,x})]_x
\\&+\frac{1}{2}m_{12}(u_{11,x}u_{21}-u_{11}u_{21,x})-\frac{1}{2}m_{21}(u_{11}u_{12,x}-u_{11,x}u_{12}),
\\
m_{12,t}=&\frac{1}{2}[m_{12}(u_{11}^2-u_{11,x}^2+u_{12}u_{21}-u_{12,x}u_{21,x})]_x
\\&+m_{11}(u_{11}u_{12,x}-u_{11,x}u_{12})+\frac{1}{2}m_{12}(u_{12,x}u_{21}-u_{12}u_{21,x}),
\\
m_{21,t}=&\frac{1}{2}[m_{21}(u_{11}^2-u_{11,x}^2+u_{12}u_{21}-u_{12,x}u_{21,x})]_x
\\&+m_{11}(u_{11}u_{21,x}-u_{11,x}u_{21})+\frac{1}{2}m_{21}(u_{12}u_{21,x}-u_{12,x}u_{21}),
\\
m_{11}=&u_{11}-u_{11,xx}, \quad m_{12}=u_{12}-u_{12,xx}, \quad m_{21}=u_{21}-u_{21,xx}.
\end{split}\right. \label{ceq}
\end{eqnarray}
Apparently, this system is reduced to the CH equation (\ref{CH}) and the cubic nonlinear CH equation (\ref{mCH}) as $u_{11}=0,~u_{21}=2$ and $u_{12}=u_{21}=0$.
Therefore, it is a three-component formation based on the CH equation (\ref{CH}) and the cubic nonlinear CH equation (\ref{mCH}), and we may call equation (\ref{ceq}) the 3CH model.
We show that the 3CH system is Hamiltonian and possesses a Lax pair and infinitely many conservation laws. Furthermore, this  three-component system
admits the single peakon of traveling wave type as well as multi-peakon solutions.
Additionally, we pay attention to the reductions of the 3CH system so that a new integrable perturbed CH equation with cubic nonlinearity is generated to possess peakon solutions.

The whole paper is organized as follows. In section 2, a Lax
pair, Hamiltonian structure, and conservation laws of equation (\ref{ceq}) are presented.
In section 3, the single-peakon and multi-peakon solutions of equation (\ref{ceq}) are given.
Section 4 discusses a new integrable perturbation of cubic nonlinear CH equation obtained from the reductions of system (\ref{ceq}).
Some conclusions and open problems are described in section 5.

\section{Lax pair, Hamiltonian form and conservation laws}

Let us first introduce the $sl(2)$ valued matrices $u$ and $m$ as follows:
\begin{equation}
u=\left(\begin{array}{cc}
u_{11}& u_{12}\\
u_{21}& -u_{11}
\end{array}
\right),
\quad
m=\left(\begin{array}{cc}
u_{11}-u_{11,xx}& u_{12}-u_{12,xx}\\
u_{21}-u_{21,xx}& -u_{11}+u_{11,xx}
\end{array}
\right)
\triangleq
\left(\begin{array}{cc}
m_{11} & m_{12}\\
m_{21} & -m_{11}
\end{array}
\right).
\label{um}
\end{equation}
Using this notation, equation (\ref{ceq}) can be expressed in a nice matrix equation form
\begin{eqnarray}
m_t=\frac{1}{2}[m(u^2-u_x^2)]_x+\frac{1}{4}[m(uu_x-u_xu)-(uu_x-u_xu)m], \quad m=u-u_{xx},
\label{meq}
\end{eqnarray}
where $u$ and $m$ are two $sl(2)$ matrices (\ref{um}).

We consider a pair of linear spectral problems
\begin{eqnarray}
\phi_x=U\phi,\quad \phi_t=V\phi,
\label{LP}
\end{eqnarray}
with
\begin{eqnarray}
\begin{split}
\phi&=(\phi_1,\phi_2,\phi_3,\phi_4)^T,\\
U&=\frac{1}{2}\left( \begin{array}{cc} -I_2 & \lambda m\\
 \lambda  m &  I_2 \\ \end{array} \right)\triangleq
\left(\begin{array}{cc}
U_{11} & U_{12}\\
U_{21} & U_{22}
\end{array}
\right),
\\
V&=\frac{1}{2}\left( \begin{array}{cc} \lambda^{-2}I_2-\frac{1}{2}(u^2-u_x^2+uu_x-u_xu) & -\lambda^{-1}(u-u_x)+\frac{1}{2}\lambda m(u^2-u_x^2)
\\ -\lambda^{-1}(u+u_x)+\frac{1}{2}\lambda m(u^2-u_x^2) & -\lambda^{-2}I_2+\frac{1}{2}(u^2-u_x^2+u_xu-uu_x) \\ \end{array} \right)
\\&\triangleq
\left(\begin{array}{cc}
V_{11} & V_{12}\\
V_{21} & V_{22}
\end{array}
\right),
\label{UV}
\end{split}
\end{eqnarray}
where $\lambda$ is a spectral parameter, $I_2$ is the $2\times 2$ identity matrix, and $u$ and $m$ are $sl(2)$ valued matrices (\ref{um}).

The compatibility condition of (\ref{LP}) generates
\begin{eqnarray}
U_t-V_x+[U,V]=0.\label{cc}
\end{eqnarray}
Substituting the expressions of $U$ and $V$ given by (\ref{UV}) into (\ref{cc}), we find that (\ref{cc}) is nothing
but the matrix equation (\ref{meq}). Hence, (\ref{LP}) exactly gives the Lax pair of equation (\ref{ceq}).

Let
\begin{eqnarray}
\begin{split}
J&=\left( \begin{array}
{ccc} J_{11} &  J_{12} & J_{13} \\  J_{21} &  J_{22} & J_{23} \\ J_{31} &  J_{32} & J_{33}\\
\end{array} \right),
\label{JK}
\end{split}
\end{eqnarray}
where
\begin{eqnarray}
\begin{split}
J_{11}=& \partial m_{11}\partial^{-1}m_{11}\partial+\frac{1}{2}m_{12}\partial^{-1}m_{21}+\frac{1}{2}m_{21}\partial^{-1}m_{12},
\\ J_{12}=& \partial m_{11}\partial^{-1} m_{12}\partial-m_{12}\partial^{-1} m_{11},
\\ J_{13}=& \partial m_{11}\partial^{-1} m_{21}\partial-m_{21}\partial^{-1} m_{11},
\\ J_{21}=&-J_{12}^{\ast}= \partial m_{12}\partial^{-1} m_{11}\partial-m_{11}\partial^{-1} m_{12},
\\ J_{22}=& \partial m_{12}\partial^{-1} m_{12}\partial-m_{12}\partial^{-1} m_{12},
\\ J_{23}=& \partial m_{12}\partial^{-1} m_{21}\partial+2 m_{11}\partial^{-1} m_{11}+m_{12}\partial^{-1} m_{21},
\\ J_{31}=&-J_{13}^{\ast}= \partial m_{21}\partial^{-1} m_{11}\partial-m_{11}\partial^{-1} m_{21},
\\ J_{32}=&-J_{23}^{\ast}= \partial m_{21}\partial^{-1} m_{12}\partial+2 m_{11}\partial^{-1} m_{11}+m_{21}\partial^{-1} m_{12},
\\ J_{33}=& \partial m_{21}\partial^{-1} m_{21}\partial-m_{21}\partial^{-1} m_{21}.
\end{split}
\end{eqnarray}
It is easy to check $J$ is skew-symmetric. By direct but tedious calculations, we can prove the Jacobi identity
\begin{eqnarray}
\langle \alpha, J'[J\beta]\gamma\rangle+\langle \beta, J'[J\gamma]\alpha\rangle+\langle \gamma, J'[J\alpha]\beta\rangle=0,
\label{Jacb}
\end{eqnarray}
where
\begin{eqnarray}
\alpha=(\alpha_1,\alpha_2,\alpha_3)^T,\quad \beta=(\beta_1,\beta_2,\beta_3)^T, \quad \gamma=(\gamma_1,\gamma_2,\gamma_3)^T.
\label{alpha}
\end{eqnarray}
Thus $J$ is Hamiltonian operator.

\begin{proposition}
Equation (\ref{ceq}) can be rewritten as a Hamiltonian form
\begin{eqnarray}
\left(m_{11,t},~ m_{12,t},~m_{21,t}\right)^{T}=J \left(\frac{\delta H}{\delta m_{11}},~\frac{\delta H}{\delta  m_{12}},~\frac{\delta H}{\delta m_{21}}\right)^{T},\label{BH}
\end{eqnarray}
where $J$ is given by (\ref{JK}), and
\begin{eqnarray}
\begin{split}
H_1&=\frac{1}{2}\int_{-\infty}^{+\infty}(u_{11}^2+u_{12}u_{21}+u_{11,x}^2+u_{12,x}u_{21,x})dx.
\end{split}
\label{H}
\end{eqnarray}
\end{proposition}

Next let us construct the conservation laws of equation (\ref{ceq}) with the method developed in \cite{WSK,TW}.
We consider
\begin{eqnarray}
\left( \begin{array}{c} \Phi_1\\ \Phi_2 \\ \end{array} \right)_x=
\left(\begin{array}{cc}
U_{11} & U_{12}\\
U_{21} & U_{22}
\end{array}
\right)
\left( \begin{array}{c} \Phi_1\\ \Phi_2 \\ \end{array} \right),
\quad
\left( \begin{array}{c} \Phi_1\\ \Phi_2 \\ \end{array} \right)_t=
\left(\begin{array}{cc}
V_{11} & V_{12}\\
V_{21} & V_{22}
\end{array}
\right)
\left( \begin{array}{c} \Phi_1\\ \Phi_2 \\ \end{array} \right),
\label{LP2}
\end{eqnarray}
where
$\Phi_1$, $\Phi_2$, $U_{ij}$ and $V_{ij}$, $1\leq i, j\leq 2$, are all $2\times2$ matrices. 
Let $\Omega=\Phi_2\Phi_1^{-1}$, then we may check that $\Omega$ satisfies the following matrix Riccati equation
\begin{eqnarray}
\Omega_x=U_{21}+U_{22}\Omega-\Omega U_{11}-\Omega U_{12}\Omega.
\label{ric}
\end{eqnarray}
From the compatibility condition of (\ref{LP2}), we arrive at the conservation law
\begin{eqnarray}
\left[tr(U_{11}+U_{12}\Omega)]_t=[tr(V_{11}+V_{12}\Omega)\right]_x,
\label{CL}
\end{eqnarray}
where $tr(A)$ denotes the trace of a matrix $A$.

Substituting the expressions $U_{ij}$ and $V_{ij}$, $1\leq i, j\leq 2$, given by (\ref{UV}) into (\ref{ric}) and (\ref{CL}),  we immediately obtain the Riccati equation and conservation law for our equation (\ref{ceq})
\begin{eqnarray}
&&\Omega_x=\frac{1}{2}\lambda m+\Omega-\frac{1}{2}\lambda\Omega m\Omega, \label{ric2}
\\
&&[tr(m\Omega)]_t=\left[tr\left(-\lambda^{-2}(u-u_x)\Omega-\frac{1}{2}\lambda^{-1}(u^2-u_x^2)+\frac{1}{2}m(u^2-u_x^2)\Omega\right)\right]_x.
\label{CL2}
\end{eqnarray}
Equation (\ref{CL2}) shows that $tr(m\Omega)$ is a generating function of the conserved densities. To derive the explicit forms of conserved densities,
we expand $m\Omega$ in terms of negative powers of $\lambda$ as below:
\begin{equation}
m\Omega=\sum_{j=0}^{\infty}\omega_j\lambda^{-j}.\label{oe1}
\end{equation}
Substituting (\ref{oe1}) into (\ref{ric2}) and equating the coefficients of powers of $\lambda$, we arrive at
\begin{eqnarray}
\begin{split}
\omega_{0}&=(m_{11}^2+m_{12}m_{21})^{\frac{1}{2}}I_2, \qquad \omega_{1}=\omega_{0}^{-1}[\omega_{0}-m(m^{-1}\omega_{0})_x],
\\
\omega_{j+1}&=\omega_{0}^{-1}\left[\omega_j-\frac{1}{2}\sum_{i+k=j+1,~1\leq i,k\leq j}\omega_i\omega_k-m(m^{-1}\omega_{j})_x\right],\quad j\geq 1.
\end{split}
\label{wj}
\end{eqnarray}
Inserting (\ref{oe1}) and (\ref{wj}) into (\ref{CL2}), we finally obtain the following infinitely many conserved densities $\rho_j$
and the associated fluxes $F_j$:
\begin{eqnarray}
\begin{split}
\rho_{0}&=tr(\omega_{0})=2(m_{11}^2+m_{12}m_{21})^{\frac{1}{2}},
\\
F_0&=\frac{1}{2}tr[(u^2-u_x^2)\omega_{0}]=(u_{11}^2-u_{11,x}^2+u_{12}u_{21}-u_{12,x}u_{21,x})(m_{11}^2+m_{12}m_{21})^{\frac{1}{2}},
\\
\rho_{1}&=tr(\omega_{1}), ~~ F_1=\frac{1}{2}tr[-(u^2-u_x^2)+(u^2-u_x^2)\omega_{1}],
\\
\rho_{2}&=tr(\omega_{2}), ~~ F_2=tr[-(u-u_x)m^{-1}\omega_0+\frac{1}{2}(u^2-u_x^2)\omega_{2}],
\\
\rho_{j+1}&=tr(\omega_{j+1}), ~~F_{j+1}=tr[-(u-u_x)m^{-1}\omega_{j-1}+\frac{1}{2}(u^2-u_x^2)\omega_{j+1}],\quad j\geq 2,
\end{split}
\label{rjj}
\end{eqnarray}
where $\omega_j$ is given by (\ref{wj}).

\section{Peakon solutions}

Let us suppose the single peakon solution of (\ref{ceq}) as the following form
\begin{eqnarray}
u_{11}=c_{11}e^{-\mid x-ct\mid},\quad u_{12}=c_{12}e^{-\mid x-ct\mid}, \quad u_{21}=c_{21}e^{-\mid x-ct\mid}, \label{ocp}
\end{eqnarray}
where the constants $c_{11}$, $c_{12}$ and $c_{21}$ are to be determined.
The first order derivatives of $u_{11}$,$u_{12}$ and $u_{21}$
do not exist at $x=ct$. Thus (\ref{ocp}) can not be a solution of equation (\ref{ceq}) in the classical sense. However, with the help of distribution theory we have
\begin{eqnarray}
\begin{split}
u_{11,x}&=-c_{11}sgn(x-ct)e^{-\mid x-ct\mid}, \quad m_{11}=2c_{11}\delta(x-ct),
\\
u_{12,x}&=-c_{12}sgn(x-ct)e^{-\mid x-ct\mid}, \quad m_{12}=2c_{12}\delta(x-ct),
\\
u_{21,x}&=-c_{21}sgn(x-ct)e^{-\mid x-ct\mid}, \quad m_{21}=2c_{21}\delta(x-ct).
\end{split}
\label{ocpd}
\end{eqnarray}
Substituting (\ref{ocp}) and (\ref{ocpd}) into (\ref{ceq}) and integrating in the distribution sense,
one can see that $c_{11}$, $c_{12}$ and $c_{21}$ should satisfy
\begin{eqnarray}
c_{11}^2+c_{12}c_{21}=-3c. \label{C1}
\end{eqnarray}
Thus the single peakon solution becomes
\begin{eqnarray}
u_{11}=c_{11}e^{-\mid x+\frac{c_{11}^2+c_{12}c_{21}}{3}t\mid},\quad u_{12}=c_{12}e^{-\mid x+\frac{c_{11}^2+c_{12}c_{21}}{3}t\mid},
\quad u_{21}=c_{21}e^{-\mid x+\frac{c_{11}^2+c_{12}c_{21}}{3}t\mid}. \label{ocp1}
\end{eqnarray}

In general, we assume $N$-peakon solution has the following form
\begin{eqnarray}
u_{11}=\sum_{j=1}^N p_j(t)e^{-\mid x-q_j(t)\mid}, ~~u_{12}=\sum_{j=1}^N r_j(t)e^{-\mid x-q_j(t)\mid},~~ u_{21}=\sum_{j=1}^N s_j(t)e^{-\mid x-q_j(t)\mid}.
\label{NP}
\end{eqnarray}
In the distribution sense, we have
\begin{eqnarray}
\begin{split}
u_{11,x}&=-\sum_{j=1}^N p_jsgn(x-q_j)e^{-\mid x-q_j\mid}, \quad m_{11}=2\sum_{j=1}^N p_j\delta(x-q_j),
\\
u_{12,x}&=-\sum_{j=1}^N r_jsgn(x-q_j)e^{-\mid x-q_j\mid}, \quad m_{12}=2\sum_{j=1}^N r_j\delta(x-q_j),
\\
u_{21,x}&=-\sum_{j=1}^N s_jsgn(x-q_j)e^{-\mid x-q_j\mid}, \quad m_{21}=2\sum_{j=1}^N s_j\delta(x-q_j).
\end{split}
\label{Npd}
\end{eqnarray}
Substituting (\ref{NP}) and (\ref{Npd}) into (\ref{ceq}) and integrating through test functions with compact support, we obtain the $N$-peakon dynamic system as follows:
\begin{eqnarray}
\left\{
\begin{split}
p_{j,t}=&\frac{1}{2}r_j\sum_{i,k=1}^N p_is_k \left(sgn(q_j-q_k)-sgn(q_j-q_i)\right)e^{ -\mid q_j-q_k\mid-\mid q_j-q_i\mid}
\\&-\frac{1}{2}s_j\sum_{i,k=1}^N p_ir_k \left(sgn(q_j-q_i)-sgn(q_j-q_k)\right)e^{ -\mid q_j-q_k\mid-\mid q_j-q_i\mid},\\
r_{j,t}=&p_j\sum_{i,k=1}^N p_ir_k \left(sgn(q_j-q_i)-sgn(q_j-q_k)\right)e^{ -\mid q_j-q_k\mid-\mid q_j-q_i\mid}
\\&+\frac{1}{2}r_j\sum_{i,k=1}^N r_is_k \left(sgn(q_j-q_k)-sgn(q_j-q_i)\right)e^{ -\mid q_j-q_k\mid-\mid q_j-q_i\mid},\\
s_{j,t}=&p_j\sum_{i,k=1}^N p_is_k \left(sgn(q_j-q_i)-sgn(q_j-q_k)\right)e^{ -\mid q_j-q_k\mid-\mid q_j-q_i\mid}
\\&-\frac{1}{2}s_j\sum_{i,k=1}^N r_is_k \left(sgn(q_j-q_k)-sgn(q_j-q_i)\right)e^{ -\mid q_j-q_k\mid-\mid q_j-q_i\mid},\\
q_{j,t}=&\frac{1}{6}(p_j^2+r_js_j)-\frac{1}{2}\sum_{i,k=1}^N(p_ip_k+r_is_k)\left(1-sgn(q_j-q_i)sgn(q_j-q_k)\right)e^{ -\mid q_j-q_i\mid-\mid q_j-q_k\mid}.
\end{split}
\right.\label{dNcp}
\end{eqnarray}
We still do not know whether this system is integrable for $N\geq 2$ under a Poisson structure.

\section{Reductions and a new integrable perturbation equation}
As mentioned above, system (\ref{ceq}) can be reduced to the CH equation (\ref{CH}) and the cubic nonlinear CH equation (\ref{mCH}) as $u_{11}=0,~u_{21}=2$ and $u_{12}=u_{21}=0$. Now we discuss the two-component reductions of system (\ref{ceq}).

\subsection*{Example 1.~~The integrable two-component system proposed in \cite{XQ}}
As $u_{11}=0$, equation (\ref{ceq}) is reduced to the two-component equation
\begin{eqnarray}
\left\{\begin{array}{l}
m_{12,t}=\frac{1}{2}[m_{12}(u_{12}u_{21}-u_{12,x}u_{21,x})]_x+\frac{1}{2}m_{12}(u_{12,x}u_{21}-u_{12}u_{21,x}),
\\
m_{21,t}=\frac{1}{2}[m_{21}(u_{12}u_{21}-u_{12,x}u_{21,x})]_x+\frac{1}{2}m_{21}(u_{12}u_{21,x}-u_{12,x}u_{21}),
\\ m_{12}=u_{12}-u_{12,xx},
\\ m_{21}=u_{21}-u_{21,xx},
\end{array}\right. \label{teq1}
\end{eqnarray}
which is exactly the system we derived in \cite{XQ}. For the bi-Hamiltonian structure and peakon solutions of this system, one may see \cite{XQ}.

\subsection*{Example 2.~~The integrable two-component system presented in \cite{QSY}}
As $u_{12}=u_{21}$, equation (\ref{ceq}) is reduced to a two-component equation
\begin{eqnarray}
\left\{\begin{array}{l}
m_{11,t}=\frac{1}{2}[m_{11}(u_{11}^2+u_{12}^2-u_{11,x}^2-u_{12,x}^2)]_x+m_{12}(u_{11,x}u_{12}-u_{11}u_{12,x}),
\\
m_{12,t}=\frac{1}{2}[m_{12}(u_{11}^2+u_{12}^2-u_{11,x}^2-u_{12,x}^2)]_x+m_{11}(u_{11}u_{12,x}-u_{11,x}u_{12}),
\\ m_{11}=u_{11}-u_{11,xx},
\\ m_{12}=u_{12}-u_{12,xx},
\end{array}\right. \label{teq2}
\end{eqnarray}
which was proposed by Qu, Song and Yao in \cite{QSY}. Here in our paper, we want to derive the peakon solutions to this system. Suppose $N$-peakon solution of (\ref{teq2}) as the form
\begin{eqnarray}
u_{11}=\sum_{j=1}^N p_j(t)e^{-\mid x-q_j(t)\mid}, ~~u_{12}=\sum_{j=1}^N r_j(t)e^{-\mid x-q_j(t)\mid}.
\label{NP2}
\end{eqnarray}
From (\ref{dNcp}) and the reduction condition $u_{12}=u_{21}$, we immediately arrive at the $N$-peakon dynamic system of (\ref{teq2}):
\begin{eqnarray}
\left\{
\begin{split}
p_{j,t}=&r_j\sum_{i,k=1}^N p_ir_k \left(sgn(q_j-q_k)-sgn(q_j-q_i)\right)e^{ -\mid q_j-q_k\mid-\mid q_j-q_i\mid},\\
r_{j,t}=&p_j\sum_{i,k=1}^N p_ir_k \left(sgn(q_j-q_i)-sgn(q_j-q_k)\right)e^{ -\mid q_j-q_k\mid-\mid q_j-q_i\mid},\\
q_{j,t}=&\frac{1}{6}(p_j^2+r_j^2)-\frac{1}{2}\sum_{i,k=1}^N(p_ip_k+r_ir_k)\left(1-sgn(q_j-q_i)sgn(q_j-q_k)\right)e^{ -\mid q_j-q_i\mid-\mid q_j-q_k\mid}.
\end{split}
\right.\label{dNcp2}
\end{eqnarray}
For $N=1$, we find the single-peakon solution reads
\begin{eqnarray}
u_{11}=c_1e^{-\mid x+\frac{c_1^2+c_2^2}{3}t\mid},\quad u_{12}=c_{2}e^{-\mid x+\frac{c_1^2+c_2^2}{3}t\mid},\label{ocpnp2}
\end{eqnarray}
where $c_{1}$ and $c_{2}$ are integration constants.

\begin{figure}
\begin{minipage}[t]{0.5\linewidth}
\centering
\includegraphics[width=2.2in]{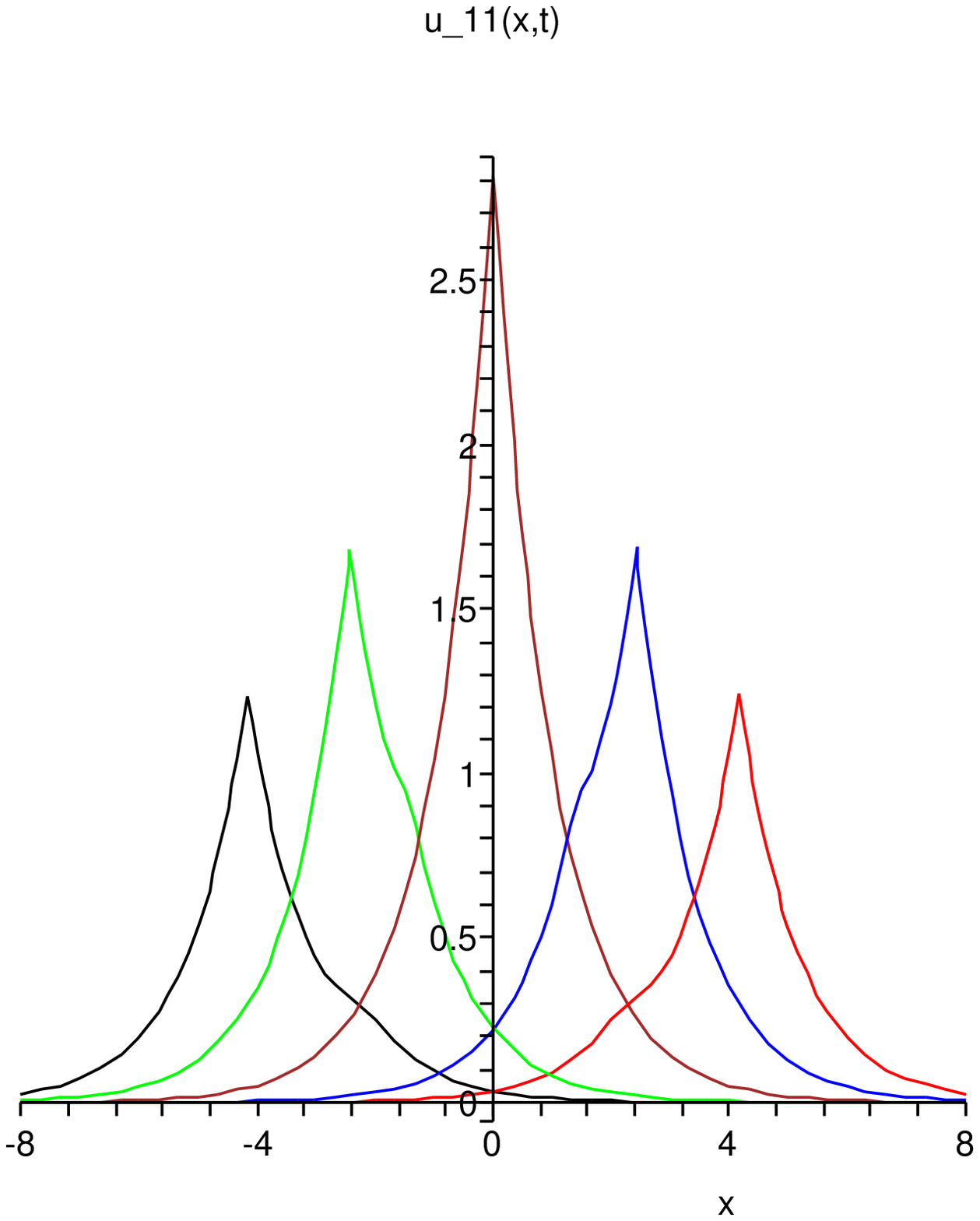}
\caption{\small{ The two-peakon dynamic for the potential $u_{11}(x,t)$ in (\ref{2uv}). Red line: $t=-2$; Blue line: $t=-1$; Brown line: $t=0$ (collision); Green line: $t=1$; Black line: $t=2$. }}
\label{F1u}
\end{minipage}
\hspace{2.0ex}
\begin{minipage}[t]{0.5\linewidth}
\centering
\includegraphics[width=2.2in]{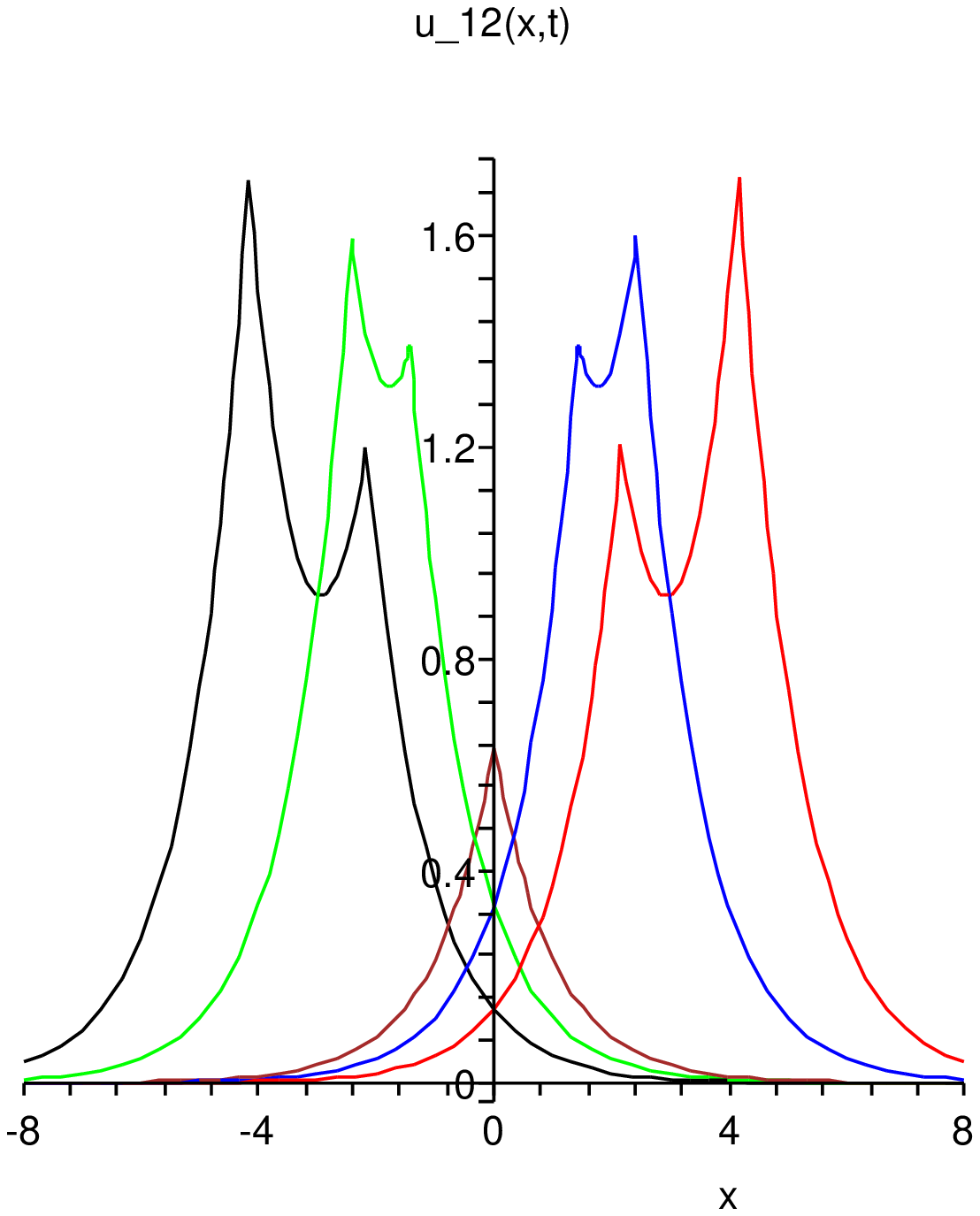}
\caption{\small{ The two-peakon dynamic for the potential $u_{12}(x,t)$ in (\ref{2uv}). Red line: $t=-2$; Blue line: $t=-1$; Brown line: $t=0$ (collision); Green line: $t=1$; Black line: $t=2$. }}
\label{F1v}
\end{minipage}
\end{figure}

For $N=2$,
we may solve (\ref{dNcp2}) as
\begin{eqnarray}
\left\{\begin{array}{l}
q_{1}(t)=-\frac{1}{3}A_1^2t+\frac{3A_1A_2\cos(A_3-A_4)}{|A_1^2-A_2^2|}sgn(t)\left(e^{-\frac{1}{3}\mid(A_1^2-A_2^2) t\mid}-1\right),\\
q_{2}(t)=-\frac{1}{3}A_2^2t+\frac{3A_1A_2\cos(A_3-A_4)}{|A_1^2-A_2^2|}sgn(t)\left(e^{-\frac{1}{3}\mid(A_1^2-A_2^2) t\mid}-1\right),\\
p_{1}(t)=A_1\sin(\frac{3A_1A_2\sin(A_3-A_4)}{A_1^2-A_2^2}e^{-\frac{1}{3}\mid(A_1^2-A_2^2) t\mid}+A_3),\\
p_{2}(t)=A_2\sin(\frac{3A_1A_2\sin(A_3-A_4)}{A_1^2-A_2^2}e^{-\frac{1}{3}\mid(A_1^2-A_2^2) t\mid}+A_4),\\
r_{1}(t)=A_1\cos(\frac{3A_1A_2\sin(A_3-A_4)}{A_1^2-A_2^2}e^{-\frac{1}{3}\mid(A_1^2-A_2^2) t\mid}+A_3),\\
r_{2}(t)=A_2\cos(\frac{3A_1A_2\sin(A_3-A_4)}{A_1^2-A_2^2}e^{-\frac{1}{3}\mid(A_1^2-A_2^2) t\mid}+A_4),
\end{array}\right. \label{2pq}
\end{eqnarray}
where $A_1$, $\cdots$, $A_4$ are integration constants. It is interesting that the amplitudes are periodic. In particular, letting $A_1=1$, $A_2=2$, $A_3=0$ and $A_4=\frac{\pi}{6}$, we obtain the two-peakon solution
of (\ref{teq2})
\begin{eqnarray}
\left\{\begin{array}{l}
u_{11}=\sin(e^{-\mid t\mid})e^{-\mid x-q_1(t)\mid}+2\sin(e^{-\mid t\mid}+\frac{\pi}{6})e^{-\mid x-q_2(t)\mid},\\
u_{12}=\cos(e^{-\mid t\mid})e^{-\mid x-q_1(t)\mid}+2\cos(e^{-\mid t\mid}+\frac{\pi}{6})e^{-\mid x-q_2(t)\mid},
\end{array}\right.
\label{2uv}
\end{eqnarray}
with
\begin{eqnarray}
q_{1}(t)=-\frac{1}{3}t+\sqrt{3}sgn(t)\left(e^{-\mid t\mid}-1\right),\quad
q_{2}(t)=-\frac{4}{3}t+\sqrt{3}sgn(t)\left(e^{-\mid t\mid}-1\right).
\label{qs1}
\end{eqnarray}
The two-peakon collides at $t=0$, since $q_1(0)=q_2(0)=0$.
See Figures \ref{F1u} and \ref{F1v} for the two-peakon dynamics of the potentials $u_{11}(x,t)$ and $u_{12}(x,t)$.

\subsection*{Example 3.~~A new integrable perturbation of the cubic nonlinear CH equation}
As $u_{12}=0$, equation (\ref{ceq}) is cast into
\begin{eqnarray}
\left\{\begin{array}{l}
m_{11,t}=\frac{1}{2}[m_{11}(u_{11}^2-u_{11,x}^2)]_x,
\\
m_{21,t}=\frac{1}{2}[m_{21}(u_{11}^2-u_{11,x}^2)]_x+m_{11}(u_{11}u_{21,x}-u_{11,x}u_{21}),
\\
m_{11}=u_{11}-u_{11,xx},
\\
m_{21}=u_{21}-u_{21,xx}.
\end{array}\right.
\label{teq3}
\end{eqnarray}
This equation is different from the standard perturbation of the cubic nonlinear CH equation
\begin{eqnarray}
\left\{\begin{array}{l}
m_{11,t}=\frac{1}{2}[m_{11}(u_{11}^2-u_{11,x}^2)]_x,
\\
m_{21,t}=\frac{1}{2}[m_{21}(u_{11}^2-u_{11,x}^2)]_x+[m_{11}(u_{11}u_{21}-u_{11,x}u_{21,x})]_x,
\\
m_{11}=u_{11}-u_{11,xx},
\\
m_{21}=u_{21}-u_{21,xx}.
\end{array}\right.
\label{teq4}
\end{eqnarray}
Thus (\ref{teq3}) is a new integrable perturbation of the cubic nonlinear CH equation.
We remark that by direct calculations one may see the second potential $u_{21}$ of the standard perturbation equation (\ref{teq4}) does not admit the peakon solution in the form of $u_{21}=p(t)e^{-\mid x-q(t)\mid}$. However, we find that our new perturbation equation (\ref{teq3}) admits peakon solutions.
In fact, suppose $N$-peakon solution of (\ref{teq3}) as the form
\begin{eqnarray}
u_{11}=\sum_{j=1}^N p_j(t)e^{-\mid x-q_j(t)\mid},~~ u_{21}=\sum_{j=1}^N s_j(t)e^{-\mid x-q_j(t)\mid},
\label{NP3}
\end{eqnarray}
 we obtain the $N$-peakon dynamic system of (\ref{teq3}) as follows:
\begin{eqnarray}
\left\{
\begin{split}
p_{j,t}=&0,\\
q_{j,t}=&\frac{1}{6}p_j^2-\frac{1}{2}\sum_{i,k=1}^N p_ip_k\left(1-sgn(q_j-q_i)sgn(q_j-q_k)\right)e^{ -\mid q_j-q_i\mid-\mid q_j-q_k\mid},\\
s_{j,t}=&p_j\sum_{i,k=1}^N p_is_k \left(sgn(q_j-q_i)-sgn(q_j-q_k)\right)e^{ -\mid q_j-q_k\mid-\mid q_j-q_i\mid}.
\end{split}
\right.\label{dNcpnp}
\end{eqnarray}
For $N=1$, we find the single-peakon solution takes the form
\begin{eqnarray}
u_{11}=\sqrt{-3c}e^{-\mid x-ct\mid},\quad u_{21}=c_{21}e^{-\mid x-ct\mid},\label{ocpnp}
\end{eqnarray}
where $c_{21}$ is an arbitrary constant.

\begin{figure}
\begin{minipage}[t]{0.5\linewidth}
\centering
\includegraphics[width=2.2in]{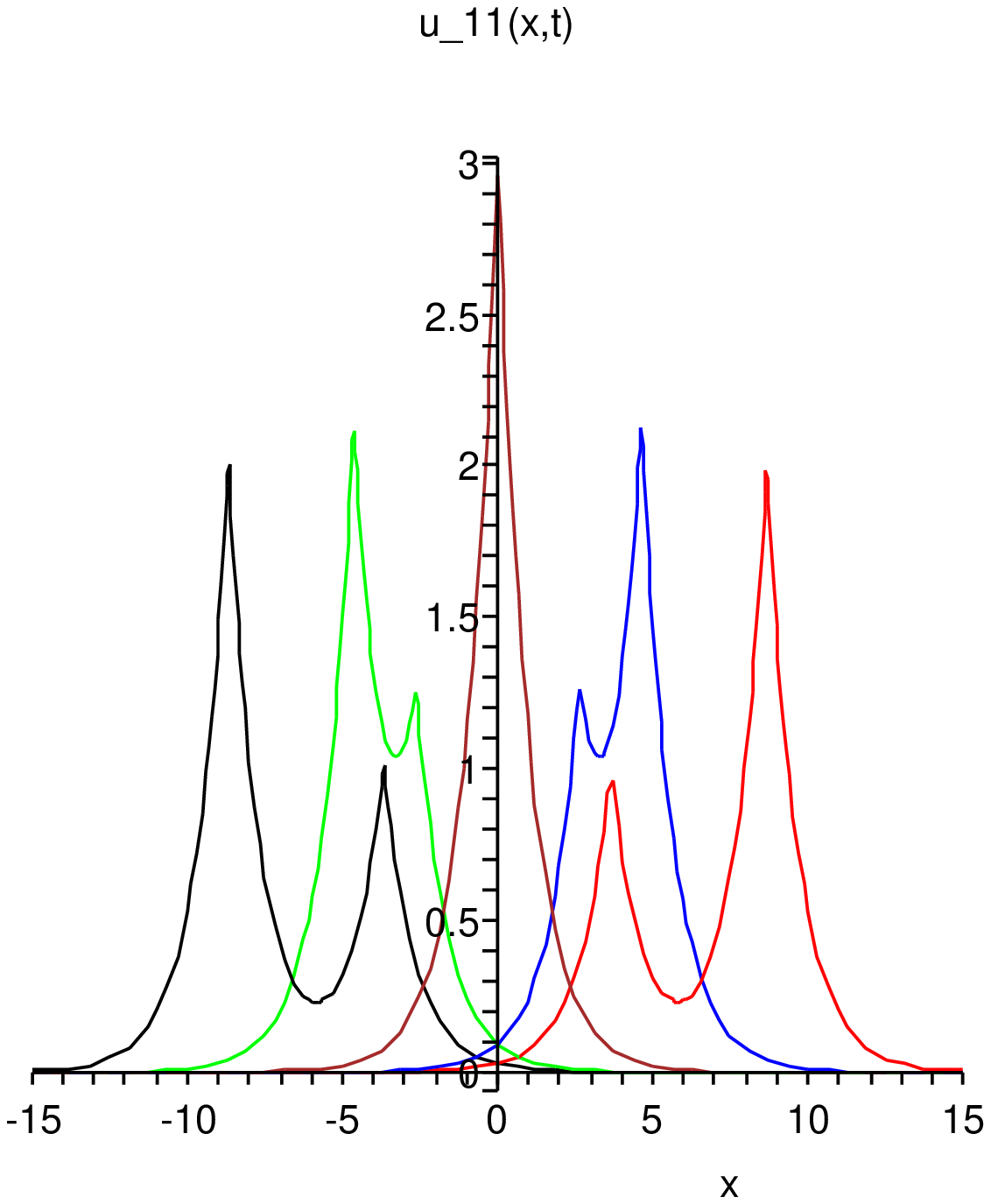}
\caption{\small{ The two-peakon dynamic for the potential $u_{11}(x,t)$ in (\ref{3uv}). Red line: $t=-5$; Blue line: $t=-2$; Brown line: $t=0$ (collision); Green line: $t=2$; Black line: $t=5$. }}
\label{F2u}
\end{minipage}
\hspace{2.0ex}
\begin{minipage}[t]{0.5\linewidth}
\centering
\includegraphics[width=2.2in]{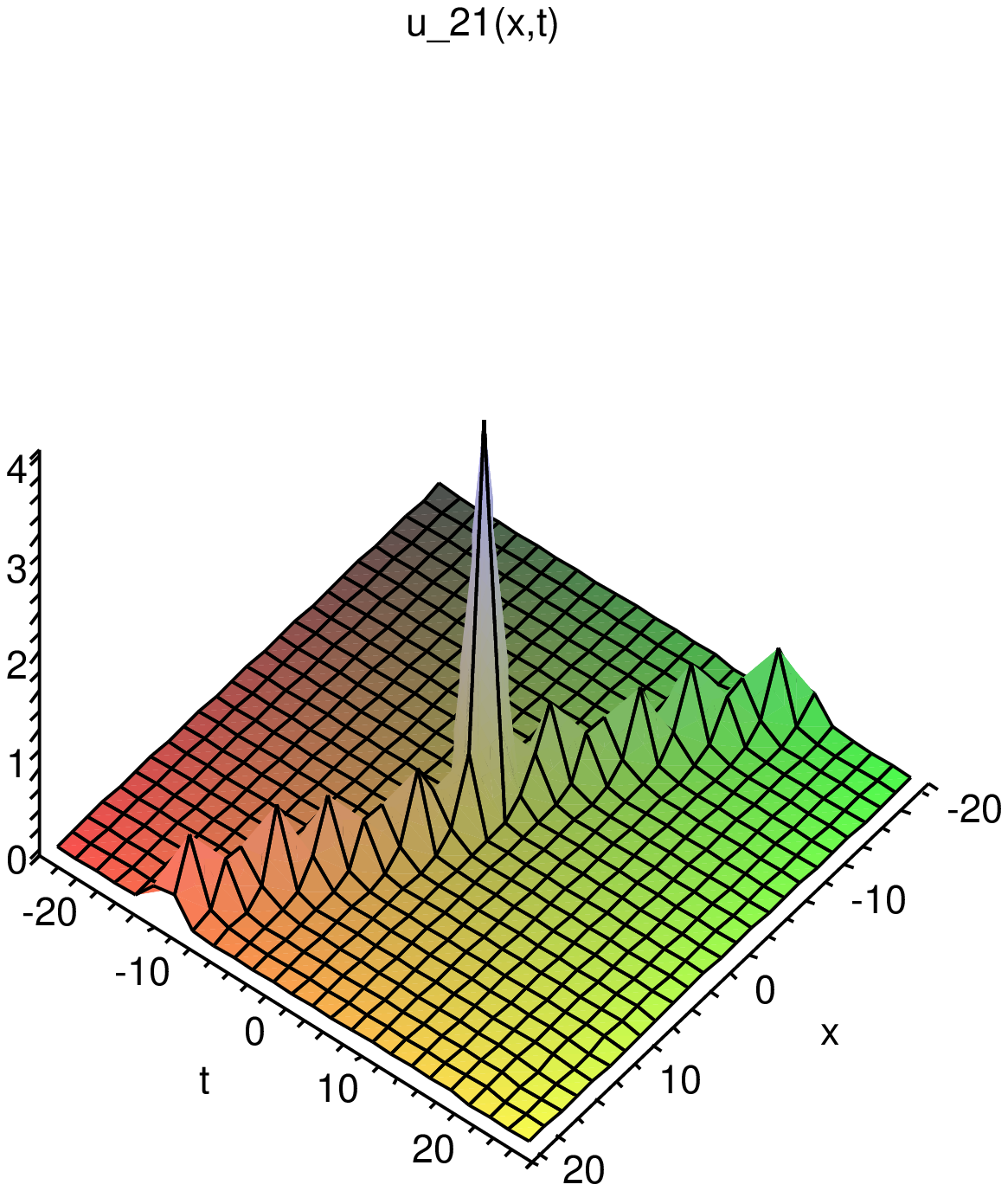}
\caption{\small{ 3-dimensional graph for the two-peakon dynamic of the potential $u_{21}(x,t)$ in (\ref{3uv}). }}
\label{F2v}
\end{minipage}
\end{figure}

For $N=2$, (\ref{dNcpnp}) becomes
\begin{eqnarray}
\left\{
\begin{split}
p_{1,t}=&p_{2,t}=0,\\
q_{1,t}=&-\frac{1}{3}p_1^2- p_1p_2e^{ -\mid q_1-q_2\mid},\\
q_{2,t}=&-\frac{1}{3}p_2^2- p_1p_2e^{ -\mid q_1-q_2\mid},\\
s_{1,t}=&p_1(p_2s_1-p_1s_2)sgn(q_1-q_2)e^{ -\mid q_1-q_2\mid},\\
s_{2,t}=&p_2(p_2s_1-p_1s_2)sgn(q_1-q_2)e^{ -\mid q_1-q_2\mid}.
\end{split}
\right.\label{dtcpnp}
\end{eqnarray}
From the first equation of (\ref{dtcpnp}), we obtain
\begin{eqnarray}
p_{1}=A_1,~~p_{2}=A_2,
\label{p}
\end{eqnarray}
where $A_1$ and $A_2$ are integration constants. Let us set $0<A_1<A_2$. Then from (\ref{dtcpnp}), we arrive at
\begin{eqnarray}
\left\{\begin{array}{l}
q_{1}(t)=-\frac{1}{3}A_1^2t+\frac{3A_1A_2}{A_2^2-A_1^2}sgn(t)\left(e^{-\frac{1}{3}(A_2^2-A_1^2)\mid t\mid}-1\right),\\
q_{2}(t)=-\frac{1}{3}A_2^2t+\frac{3A_1A_2}{A_2^2-A_1^2}sgn(t)\left(e^{-\frac{1}{3}(A_2^2-A_1^2)\mid t\mid}-1\right),\\
s_{1}(t)=\frac{3A_1A_3}{A_1^2-A_2^2}e^{-\frac{1}{3}(A_2^2-A_1^2)\mid t\mid}+A_4,\\
s_{2}(t)=\frac{1}{A_1}(A_2s_1-A_3),
\end{array}\right. \label{qepq}
\end{eqnarray}
where $A_3$ and $A_4$ are integration constants. In particular, taking $A_1=-A_3=1$, $A_2=2$ and $A_4=0$, we obtain the two-peakon solution
of (\ref{teq3})
\begin{eqnarray}
u_{11}=e^{-\mid x-q_1(t)\mid}+2e^{-\mid x-q_2(t)\mid},~~ u_{21}=e^{-\mid t\mid}e^{-\mid x-q_1(t)\mid}+(2e^{-\mid t\mid}+1)e^{-\mid x-q_2(t)\mid},
\label{3uv}
\end{eqnarray}
with
\begin{eqnarray}
q_{1}(t)=-\frac{1}{3}t+2sgn(t)\left(e^{-\mid t\mid}-1\right),\quad
q_{2}(t)=-\frac{4}{3}t+2sgn(t)\left(e^{-\mid t\mid}-1\right).
\label{qs2}
\end{eqnarray}
This two-peakon collides at the moment of $t=0$, since $q_1(0)=q_2(0)=0$.
See Figures \ref{F2u} and \ref{F2v} for the two-peakon dynamics of the potentials $u_{11}(x,t)$ and $u_{21}(x,t)$.

\section {Conclusions and discussions}
We have presented an integrable 3CH peakon system with cubic nonlinearity.
The Lax representation, Hamiltonian structure and infinitely many conservation laws of this system are investigated. We also discuss the reductions of this
system, in particular, by a reduction we found a new integrable perturbation of cubic nonlinear CH equation. Different from the standard perturbation of the cubic nonlinear CH equation, this new integrable perturbation of cubic nonlinear CH equation admits peakon solutions.

\section*{ACKNOWLEDGMENTS}

This work was partially supported by National Natural Science Foundation of China (Grant Nos. 11301229 and 11271168) and
the Natural Science Foundation of the Jiangsu Higher Education Institutions of China (Grant No. 13KJB110009).

\vspace{1cm}
\small{

}
\end{document}